# Electronic properties of superconducting $Sr_4V_2Fe_2As_2O_6$ *versus* $Sr_4Sc_2Fe_2As_2O_6$

I. R. Shein,* A. L. Ivanovskii

*Institute of Solid State Chemistry, Ural Division, Russian Academy of Sciences, Pervomaiskaya St., 91, Ekaterinburg, 620990 Russia*

**A B S T R A C T**

**First principle FLAPW-GGA calculations have been performed with the purpose to understand the peculiarities of band structure and Fermi surface topology for recently discovered superconductor: $Sr_4V_2Fe_2As_2O$ – in comparison with isostructural phase $Sr_4Sc_2Fe_2As_2O_6$. Our main finding is that the replacement of Sc on vanadium leads to drastic transformation of electronic, magnetic and conductive properties of these materials: as against non-magnetic $Sr_4Sc_2Fe_2As_2O_6$ which is formed from *non-magnetic conducting* $[Fe_2As_2]$ and *insulating* $[Sr_4Sc_2O_6]$ blocks, $Sr_4V_2Fe_2As_2O_6$ consists from *non-magnetic conducting* $[Fe_2As_2]$ blocks and $[Sr_4V_2O_6]$ blocks *which exhibit magnetic half-metallic properties*.**

* Corresponding author.
*E-mail address:* shein@ihim.uran.ru (I.R. Shein).

# 1. Introduction.

After the detection of superconductivity ($T_C \sim 26K$) in fluorine-doped La-Fe arsenide oxide $LaFeAsO_{1-x}F_x$ [1] a broad group of related so-called FeAs superconductors (SCs) was discovered, reviews [2,3].

It is important to notice, that all of these materials are based on the so-called parent phases. Among them are "1111" phases based on the layered quaternary arsenide oxides with ZrCuSiAs-type structure and the related oxygen-free "1111" phases; the three-component "122" $ThCr_2Si_2$-like and "111" PbFCl-like phases. All these materials adopt similar crystal structure with alternate stacking of conducting $[Fe_2As_2]$ blocks and insulating blocks (so-called "charge reservoirs"), such as [*Ln*O], [*AF*] blocks - for "1111" phases or atomic *A* sheets for "122" or "111" phases, where *Ln* = La, Ce … .Gd, Tb, Dy and *A* = Ca, Sr, Ba or Li, Na.

Besides, the majority of above parent phases is non-superconducting and lays on the border of their structural and magnetic instability, whereas the superconductivity arises as result of their hole or electron doping. In turn, such doping is achieved usually by various atomic replacements. Among them are: (i) the replacements of *Ln*, *A* atoms inside insulating blocks on others *Ln*, *A* atoms; (ii) the replacements of oxygen atoms inside insulating blocks on others *p* non-metal atoms; (iii) the replacements of Fe atoms on magnetic (3*d*: Mn, Co, Ni) or non-magnetic (4*d*,5*d*: Ru,Rh,Ir etc) metal atoms inside $[Fe_2As_2]$ blocks; or (iv) the replacements of As atoms inside $[Fe_2As_2]$ blocks on others pnictogens. In all cases the above mentioned nature of materials (*i.e.* the alternate stacking of conducting $[Fe_2As_2]$ blocks and insulating blocks) is preserved, see [2,3].

The very recently synthesized more complex five-component phase, $Sr_4Sc_2Fe_2As_2O_6$ [4,5] (termed further as Sc42226; which is declared as a parent phase for new FeAs SCs) adopt the same stacking ordering, where the conducting $[Fe_2As_2]$ blocks alternate with the insulating perovskite-

like[$Sr_4Sc_2O_6$] blocks. Moreover, the authors [6] report the superconductivity for Ti-doped Sc42226.

Here let us note that the insertion of 3$d$ metal atoms inside insulating blocks [$Sr_4Sc_2O_6$] should be considered as unusual atomic replacement type, which for FeAs-based systems was not used earlier. Moreover, according to [7], the insulating oxygen-containing [$Sr_4Sc_2O_6$] blocks in Ti-doped Sc42226 became conducting, and this situation differs essentially from the known picture for all others FeAs superconductors.

In this context the recent discovery [8] of V-containing phase, namely $Sr_4V_2Fe_2As_2O_6$ (termed further as V42226) seems highly interesting, as the replacement of Sc (with one $d$ electron) on vanadium (with three $d$ electrons) should exert the essential influence on the properties of these materials. Indeed for V42226 the superconductivity at 37.2 K has been reported [8]; and this material present one of the first parent phases (along with LiFeAs and NaFeAs with $T_C$ ranging from 9K to 18K [2,3]) where the superconductivity appears without doping.

In this Communication we present the results of first principle FLAPW-GGA calculations of newly discovered $Sr_4V_2Fe_2As_2O_6$; to understand the peculiarities of his band structure and Fermi surface topology the results obtained are discussed in comparison with those for isostructural $Sr_4Sc_2Fe_2As_2O_6$ phase.

Our main conclusion is that the replacement of Sc on vanadium leads to drastic transformation of electronic, magnetic and conductive properties of these materials: as against non-magnetic $Sr_4Sc_2Fe_2As_2O_6$ which is formed from **non-magnetic conducting** [$Fe_2As_2$] and **insulating** [$Sr_4Sc_2O_6$] blocks, $Sr_4V_2Fe_2As_2O_6$ consists from **non-magnetic conducting** [$Fe_2As_2$] blocks and from [$Sr_4V_2O_6$] blocks **which exhibit magnetic half-metallic properties.**

Sc42226 and V42226 adopt tetragonal (space group *P4/nmm*) crystal structure with alternating stacking of antifluorite-like [$Fe_2As_2$] and perovskite-like [$Sr(V)_4Sc_2O_6$] blocks [5,6,8]. Our calculations were carried out by means of the full-potential method with mixed basis APW+lo (FLAPW) implemented in the WIEN2k suite of programs [9]. The generalized gradient approximation (GGA) to exchange-correlation potential in the PBE form [10] was used. The full-lattice optimization for both phases including the atomic positions was done. The self-consistent calculations were considered to be converged when the difference in the total energy of the crystal did not exceed 0.1 mRy and the difference in the total electronic charge did not exceed 0.001 $e$ as calculated at consecutive steps. The data obtained are in reasonable agreement with available experiments [5,6,8]. Besides, for both phases we have calculated the total energies, $E_{tot}$, for magnetic (in assumption of ferromagnetic spin ordering) and nonmagnetic (NM) states.

As the first step, let us shortly describe the phase Sc42226, see also [7]. Firstly, the ground state of this system is NM. Figure 1 shows the near-Fermi band structure of Sc42226 as calculated along the high-symmetry $k$ lines. The Fermi level ($E_F$) is crossed by low-dispersive quasi-two-dimensional (2$D$-like) bands with mainly Fe 3$d$ character; these bands form electron pockets centered at $M$ and hole pockets centered at $\Gamma$. The Fermi surface (FS, see Fig. 1) is made of five sheets, which are cylindrical-like and parallel to the $k_z$ direction. Three of them are hole-like and are centered along the $\Gamma-Z$ direction line, the two others (electronic-like) are at the corners (M).

From the total and partial densities of states (DOSs, Figure 2) we see, that the occupy bands form three main groups (A-C) in the interval from -12.0 eV up to $E_F$. The bands with high binding energies are mainly of As 4$p$ character, whereas the states, placed in the interval from -5.3 eV up to -2.1 eV, form the Fe–As and Sc–O bonds owing to hybridization of Fe 3$d$ - As 4$p$ and Sc 3$d$ - O 2$p$ states. The near-Fermi bands are mainly of Fe 3$d$ character, as well as the low-ling empty states (peak D).

Let us note that the occupied states near the Fermi level (peak C) are formed *exclusively* by states of [Fe$_2$As$_2$] blocks, whereas perovskite-like [Sr$_4$Sc$_2$O$_6$] blocks are insulating.

Sharply different results are obtained from our FLAPW-GGA calculations for V42226 phase, Figure 3. Indeed, we find that this structure with ferromagnetic arrangement of the spins on the vanadium atoms (inside [$Sr_4V_2O_6$ blocks]) is to be energetically favorable; whereas the blocks [$Fe_2As_2$] remain non-magnetic. The calculated atomic magnetic moments on vanadium are at about 1.54 $\mu_B$. Here the V spin-down states are completely empty (see below), thus the atomic magnetic moments are determined by the occupation of the spin-up states. This situation resemble the same for other family of interesting layered materials, namely for well known quaternary borocarbides *Ln*Ni$_2$B$_2$C [11,12] with coexistence of superconducting and magnetic states, which consist from magnetic [*Ln*C] and superconducting [Ni$_2$B$_2$] blocks, see reviews [13,14].

Let us note that for spin-up channel in the near-Fermi region for V42226 phase a complex multi-band picture is formed, whereas for spin-down channel the band structure in this region became very similar to the same for Sc42226, namely these bands form electron pockets centered at *M* and hole pockets centered at Γ, see Figures 1 and 3.

The spin-projected DOS for the FM configuration of V42226 phase is shown in Figure 4. Here the exchange splitting of the V 3$d$↓,↑ manifold at E$_F$ and relative shift in the position of majority and minority spin bands are clearly visible, when the V 3$d$↓ states are pushed above E$_F$, whereas the sharp occupied peak below E$_F$ is due to the occupied V 3$d$↑ bands.

This is very unexpected result, showing that the perovskite-like blocks [$Sr_4V_2O_6$] behave as magnetic half-metals (MHMs), *i.e.* the systems with the full spin polarization at the Fermi level: $P = \{N↓(E_F) - N↑(E_F)\}/\{N↑(E_F) + N↓(E_F)\} = 1$.

The obtained magnitude of the V 3$d$-like density of states at the Fermi level for V42226 is $N^{V3d}(E_F)$=2.244 states/eV; $N^{3d}(E_F)$= 0 states/eV. In result, unlike the insulating behavior of [$Sr_4Sc_2O_6$] blocks in Sc42226, for V42226 phase the [$Sr_4V_2O_6$] blocks became conducting - for spin-up channel.

On the other hand, appreciable density of states at the Fermi level is formed due to Fe 3$d$-like states ($N^{Fe3d}_{\downarrow+\uparrow}$ ($E_F$) ~ 1.201 states/eV), *i.e.* for V42226 both types of constituent blocks ([$Sr_4V_2O_6$] and [$Fe_2As_2$]) are conducting.

Our calculations predict clear differences for FSs for isostructural Sc42226 and V42226 systems (Figs. 2 and 5). Indeed in V42226 for spin-up channel two groups of sheets arise, where one of them are tube-like in shape (with some deformations) around Γ and parallel to the $k_z$ direction. It is important to note, that these concentric cylinders are formed from non-interacting states of Fe and V atoms from the different blocks: [$Fe_2As_2$] and [$Sr_4V_2O_6$]. All bands near $E_F$ are mostly 2$D$, and quite slightly 3$D$-deformed shapes of sheets of FS are due to non-parabolic regions of the bands. The connected unclosed electronic-like sheets at the Brillouin zone corners also are mostly of 2$D$ character.

Oppositely, the Fermi surface for spin-down channel comprises five well disconnected nested hole and electronic sheets, reflecting a two-dimensional dispersion of these states - also as in all others FeAs SCs [2,3].

In summary, employing FLAPW-GGA approach, we studied the electronic properties for the newly discovered tetragonal SC $Sr_4V_2Fe_2As_2O_6$ in comparison with isostructural $Sr_4Sc_2Fe_2As_2O_6$.

Our results show that the unusual (for FeAs SCs) atomic replacement type (*i.e.* the replacements of 3$d$ atoms on others $d$ atoms inside the insulating blocks) leads to the resolute change of electronic structure, as well as of magnetic and conductive properties of these systems: as against non-magnetic $Sr_4Sc_2Fe_2As_2O_6$ which is formed from non-magnetic conducting [$Fe_2As_2$] and insulating [$Sr_4Sc_2O_6$] blocks, for $Sr_4V_2Fe_2As_2O_6$ [$Fe_2As_2$] blocks retain non-magnetic conducting character, whereas [$Sr_4V_2O_6$] blocks ***exhibit magnetic half-metallic properties.*** Thus, the situation for $Sr_4V_2Fe_2As_2O_6$ differs essentially from the known picture for all others FeAs SCs, therefore the further study of the features of the mechanisms of superconductivity for this new system, as well as the search of related materials will be of great interest.

**Acknowledgements**

Financial support from the RFBR (Grant No 09-03-00946-a) is gratefully acknowledged.

**Acknowledgements**

Financial support from the RFBR (Grant No 09-03-00946-a) is gratefully acknowledged.

**FIGURES**

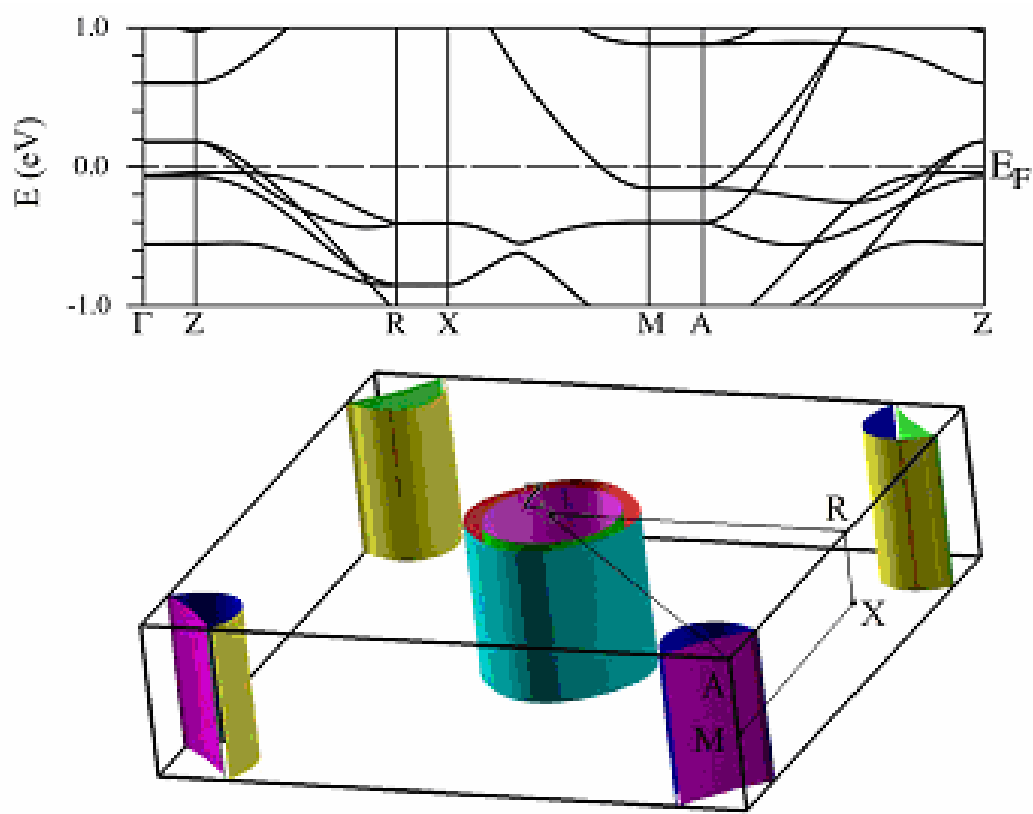

**Figure 1**. The near-Fermi electronic bands and Fermi surface for $Sr_4Sc_2Fe_2As_2O_6$.

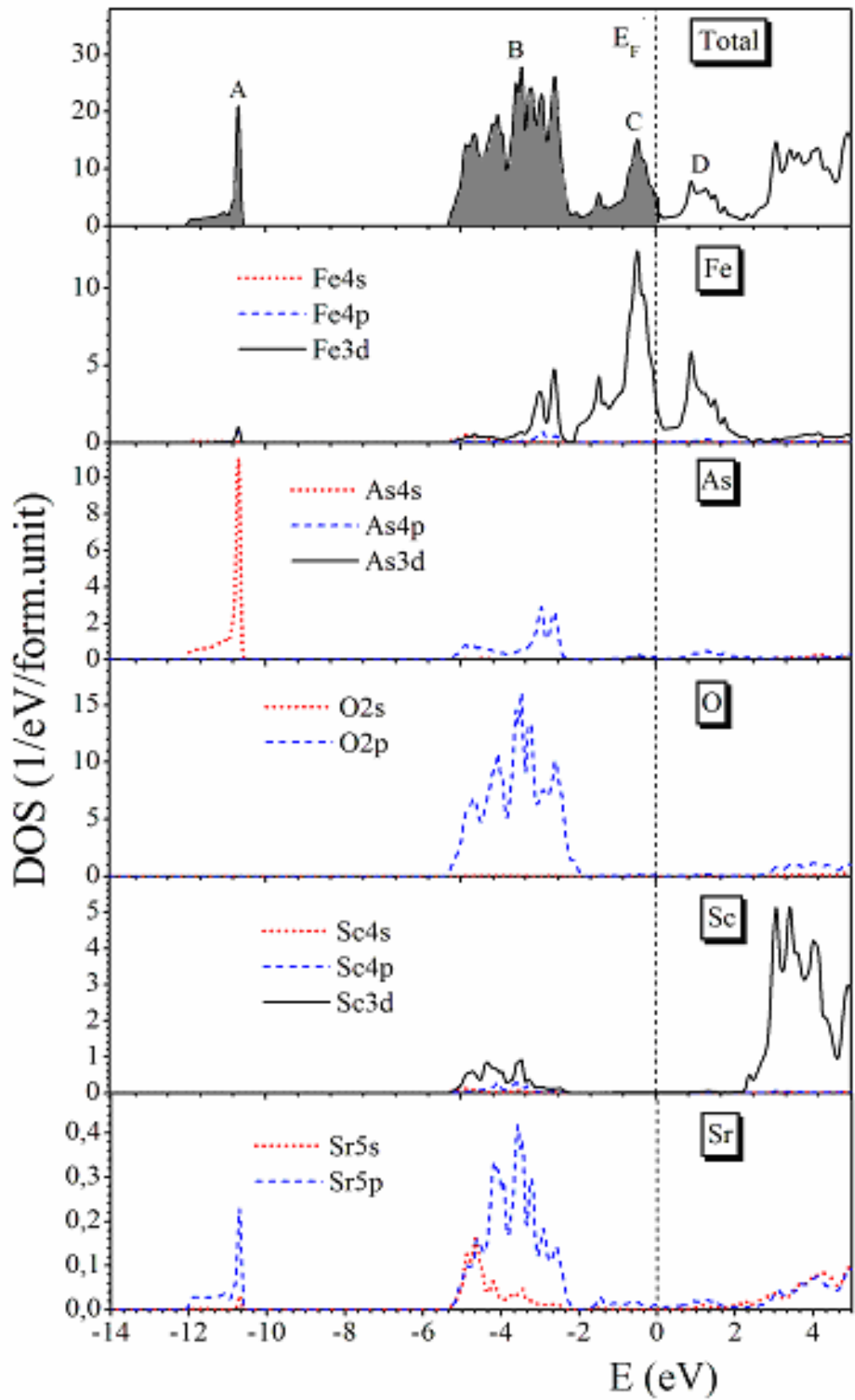

**Figure 2**. Total (*upper panel*) and partial densities of states (*bottom panels*) for $Sr_4Sc_2Fe_2As_2O_6$.

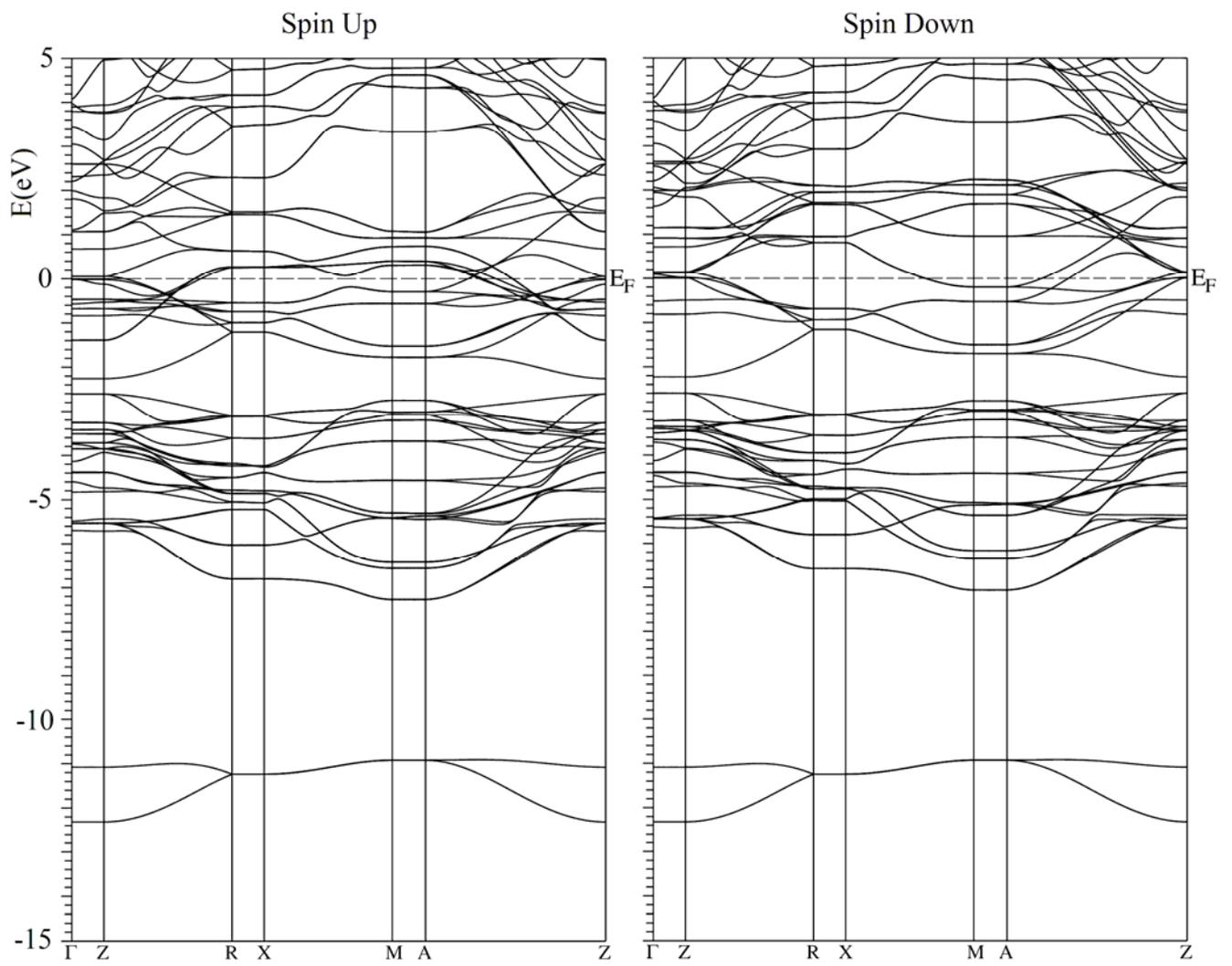

**Figure 3**. Spin-resolved electronic bands for $Sr_4V_2Fe_2As_2O_6$.

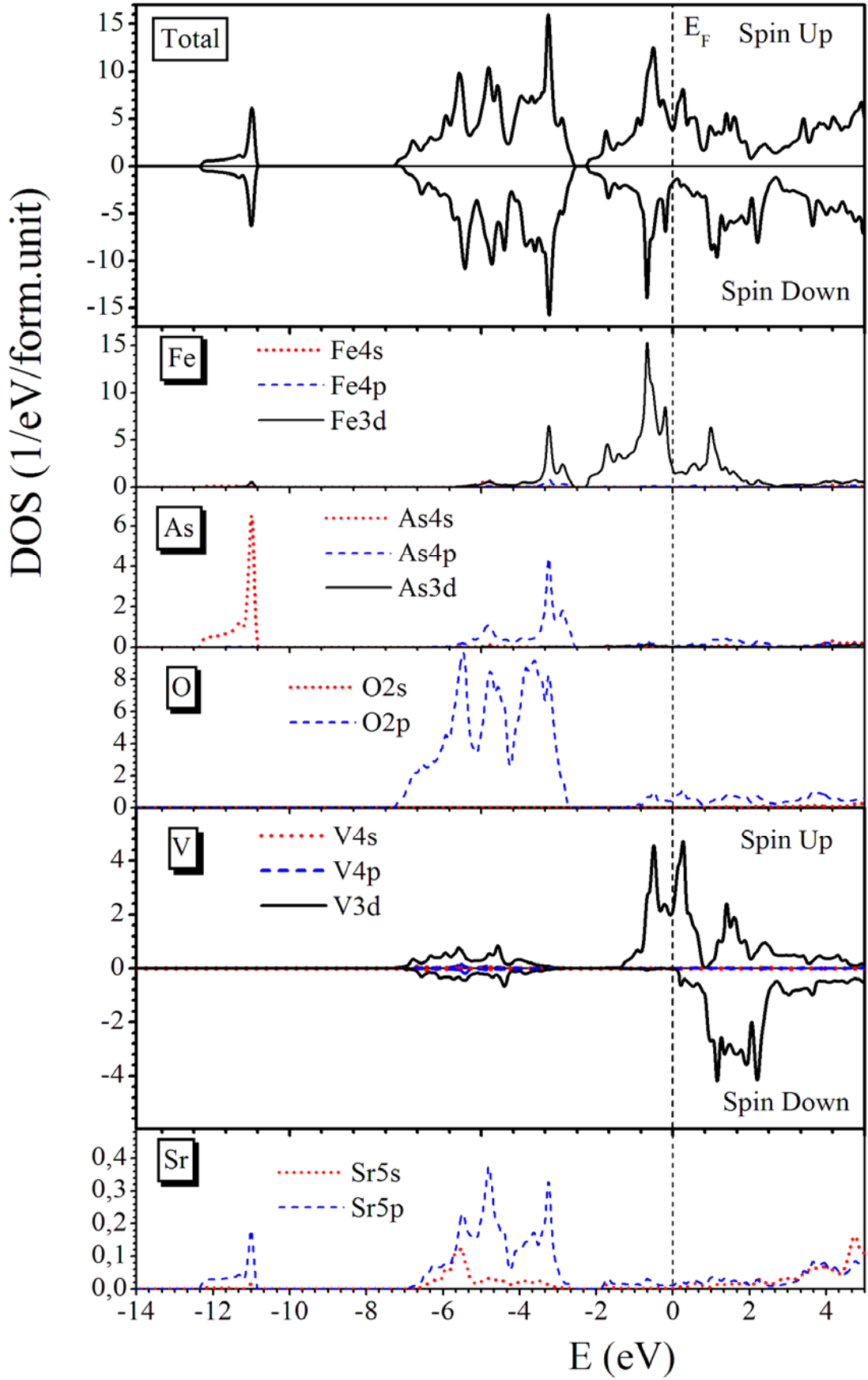

**Figure 4.** Total and partial spin-resolved densities of states for $Sr_4V_2Fe_2As_2O_6$.

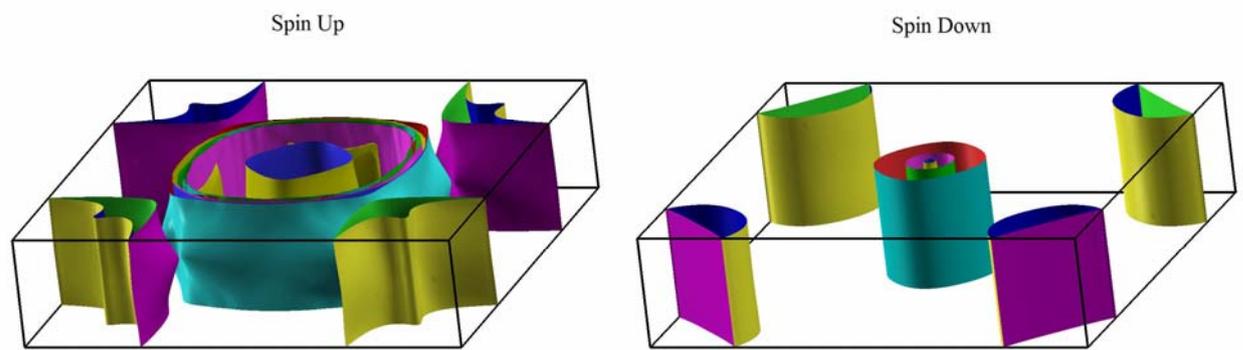

**Figure 5**. Fermi surface for $Sr_4V_2Fe_2As_2O_6$.